# Interactions between Intrinsic and Stimulus-Evoked Activity in Recurrent Neural Networks

L.F. Abbott and Kanaka Rajan

Department of Neuroscience
Department of Physiology and Cellular Biophysics
Columbia University College of Physicians and Surgeons
New York, NY 10032-2695 USA

and

Haim Sompolinsky

Racah Institute of Physics
Interdisciplinary Center for Neural Computation
Hebrew University
Jerusalem, Israel

**Introduction**

Trial-to-trial variability is an essential feature of neural responses, but its source is a subject of active debate. Response variability (Mast and Victor, 1991; Arieli et al., 1995 & 1996; Anderson et al., 2000 & 2001; Kenet et al., 2003; Petersen et al., 2003a & b; Fiser, Chiu and Weliky, 2004; MacLean et al., 2005; Yuste et al., 2005; Vincent et al., 2007) is often treated as random noise, generated either by other brain areas, or by stochastic processes within the circuitry being studied. We call such sources of variability "external" to stress the independence of this form of noise from activity driven by the stimulus. Variability can also be generated internally by the same network dynamics that generates responses to a stimulus. How can we distinguish between external and internal sources of response variability? Here we show that internal sources of variability interact nonlinearly with stimulus-induced activity, and this interaction yields a suppression of noise in the evoked state. This provides a theoretical basis and potential mechanism for the experimental observation that, in many brain areas, stimuli cause significant suppression of neuronal variability (Werner and Mountcastle, 1963; Fortier, Smith and Kalaska, 1993; Anderson et al., 2000; Friedrich and Laurent, 2004; Churchland et al., 2006; Finn, Priebe and Ferster, 2007; Mitchell, Sundberg and Reynolds, 2007; Churchland et al., 2009). The combined theoretical and experimental results suggest that internally generated activity is a significant contributor to response variability in neural circuits.

    We are interested in uncovering the relationship between intrinsic and stimulus-evoked activity in model networks and studying the selectivity of these networks to features of the stimuli driving them. The relationship between intrinsic and

extrinsically evoked activity has been studied experimentally by comparing activity patterns across cortical maps (Arieli et al., 1995 & 1996). We develop techniques for performing such comparisons in cases where there is no apparent sensory map. In addition to revealing how the temporal and spatial structure of spontaneous activity affects evoked responses, these methods can be used to infer input selectivity. Historically, selectivity was first measured by studying stimulus-driven responses (Hubel and Wiesel, 1962), and only later were similar selectivity patterns observed in spontaneous activity across the cortical surface (Arieli et al., 1995 & 1996). We argue that it is possible to work in the reverse order. Having little initial knowledge of sensory maps in our networks, we show how their spontaneous activity can inform us about the selectivity of evoked responses to input features. Throughout this study, we restrict ourselves to quantities that can be measured experimentally, such as response correlations, so our analysis methods can be applied equally to theoretical models and experimental data.

We begin by describing the network model and illustrating the types of activity it produces, using computer simulations. In particular, we illustrate and discuss a transition between two types of responses; one in which intrinsic and stimulus-evoked activity coexist, and the other in which intrinsic activity is completely suppressed. Next, we explore how the spatial patterns of spontaneous and evoked responses are related. By spatial pattern, we mean the way that activity is distributed across the different neurons of the network. Spontaneous activity is a useful indicator of recurrent effects, because it is completely determined by network feedback. Therefore, we study the impact of network connectivity on the spatial pattern of input-driven responses by comparing the spatial structure of evoked and spontaneous activity. Finally, we show how the stimulus selectivity of the network can be inferred from an analysis of its spontaneous activity.

**The Model**

Neurons in the model we consider are described by firing-rates, they do not fire individual action potentials. Such firing-rate networks are attractive because they are easier to simulate than spiking network models and are amenable to more detailed mathematical analyses. In general, as long as there is no large-scale synchronization of action potentials, firing-rate models describe network activity adequately (Shriki, Hansel and Sompolinsky, 2003; Wong and Wang, 2006). We consider a network of $N$ interconnected neurons, with neuron $i$ characterized by an activation variable $x_i$ satisfying

$$\tau \frac{dx_i}{dt} = -x_i + g \sum_{j=1}^{N} J_{ij} r_j + I_i \;.$$

The time constant $\tau$ is set to 10 ms. For all of the figures, except 5e, $N$ = 1000. The recurrent synaptic weight matrix $\mathbf{J}$ has element $J_{ij}$ describing the connection from presynaptic neuron $j$ to postsynaptic neuron $i$. Excitatory connections correspond to positive matrix elements, inhibitory connections to negative elements. The input term, $I_i$ for neuron $i$, takes various forms that will be described as we use them.

The firing rate of neuron $i$ is given by $r_i = R_0 + \phi(x_i)$ with $\phi(x) = R_0 \tanh(x/R_0)$ for $x \leq 0$ and $\phi(x) = (R_{max} - R_0)\tanh(x/(R_{max} - R_0))$ for $x > 0$. Here, $R_0$ is the background firing rate (the firing rate when $x = 0$), and $R_{max}$ is the maximum firing rate. This function allows us to specify independently the maximum firing rate, $R_{max}$, and the background rate, $R_0$, and set them to reasonable values, while retaining the general form of the commonly used tanh function. This firing rate function is plotted in Figure 1 for $R_0 = 0.1 R_{max}$, the value we use. To facilitate comparison with experimental data in a variety of systems, we report all responses relative to $R_{max}$. Similarly, we report all input currents relative to the current $I_{1/2}$ required to drive an isolated neuron to half of its maximal firing rate (see Figure 1).

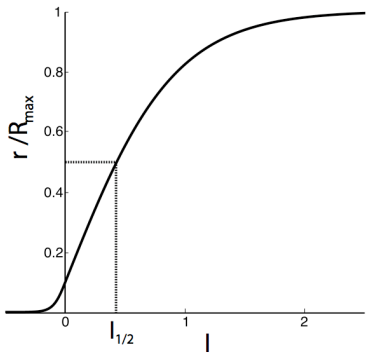

Figure 1. The firing rate function used in the network model versus the input $I$ for $R_0 = 0.1 R_{max}$ normalized by the maximum firing rate $R_{max}$. The parameter $I_{1/2}$ is defined by the dashed lines.

Although a considerable amount is known about the statistical properties of recurrent connections in cortical circuitry (Holmgren et al., 2003; Song et al., 2005), we do not have anything like the specific neuron-to-neuron wiring diagram we would need to build a truly faithful model of a cortical column or hypercolumn. Instead, we construct the connection matrix ***J*** of the model network on the basis of a statistical description of the underlying circuitry. We do this by choosing elements of the synaptic weight matrix independently and randomly from a Gaussian distribution with zero mean and variance $1/N$. We could divide the network into separate excitatory and inhibitory subpopulations, but this does not qualitatively change the network properties that we discuss (van Vreeswijk and Sompolinsky, 1996 & 1998; Rajan and Abbott, 2006).

The parameter $g$ controls the strength of the synaptic connections in the model, but because these strengths are chosen from a distribution with zero mean and nonzero variance, $g$ actually controls the size of the standard deviation of the synaptic strengths (see Discussion). Without any input ($I_i = 0$ for all $i$) and for large networks (large $N$), two spontaneous patterns of activity are seen. If $g < 1$, the network is in a trivial state in which $x_i = 0$ and $r_i = R_0$ for all neurons (all $i$). The case $g > 1$ is more interesting in that the spontaneous activity of the network is chaotic, meaning that it is irregular, non-repeating and highly sensitive to initial conditions (Sompolinsky, Crisanti, and Sommers, 1988; van Vreeswijk and Sompolinsky, 1996 & 1998). We typically use a value of $g$ = 1.5, meaning that our networks are in this chaotic state prior to activating any inputs.

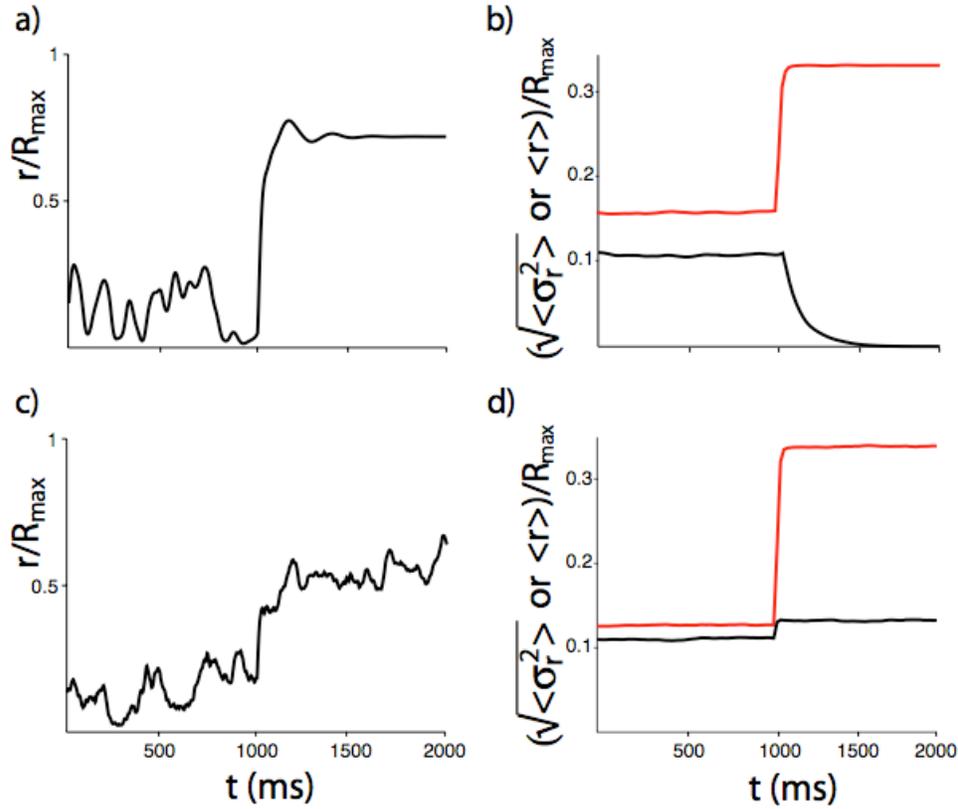

Figure 2. Firing rates and response variability normalized by the maximum firing rate, $R_{max}$, for a stimulus input stepping from zero to a constant, non-zero value at $t$ = 1000 ms. Left column shows the firing rate of a typical neuron in a network with 1000 neurons. Right column shows the average firing rate (red traces) and the square root of the average firing-rate variance (black traces) across the network neurons. a-b) A network with chaotic spontaneous activity receiving no noise input. The response variability (b, black trace) drops to zero when the stimulus input is present. c-d) A network without spontaneous activity but receiving noise input. The response variability (d, black trace) rises slightly when the step input is turned on. The stochastic input in this example was independent white-noise to each neuron, low-pass filtered with a 500 ms time constant.

**Responses to Step Input**

To begin our examination of the effects of input on chaotic spontaneous network activity, we consider the effect of a step of input (from 0 to a positive value), applied uniformly to every neuron ($I_i = I$ for all $i$). Before the input is turned on (Figure 2a & b, $t$ < 1000 ms), a typical neuron of the network shows the highly irregular activity characteristic of the chaotic spontaneous state. However, when a sufficiently strong stimulus is applied, the internally generated fluctuations are completely suppressed (Figure 2a & b, $t$ > 1000 ms). We contrast this behavior to that of external noise, by turning off the recurrent dynamics and generating fluctuations with external stochastic inputs (Figure 1c & d, $t < 1000$ ms). In this case, there is no reduction in the amplitude of the neuronal fluctuations when a stimulus is applied, in fact there is a small increase. Note that the increase in the mean activity is similar in both cases. These results reveal a critical distinction between internally and externally generated fluctuations – the former can be suppressed by a stimulus and therefore do not necessarily interfere with sensory processing.

    To reveal the nature of internally generated variability, we have considered an idealized scenario in Figure 2 in which there was no external source of noise. In reality, we expect both external and internal sources of noise to coexist in local cortical circuits. As long as the internal noise provides a substantial component of the overall variability, our qualitative results remain valid. We can simulate this situation by adding external noise (as in Figure 2c & d) to a model that exhibits chaotic spontaneous activity (as in Figure 2a & b). The result shows a sharp drop in variance at stimulus onset, but with only partial, rather than complete, suppression of response variability (Figure 3). This result is in good agreement with experimental data (Churchland et al., 2009).

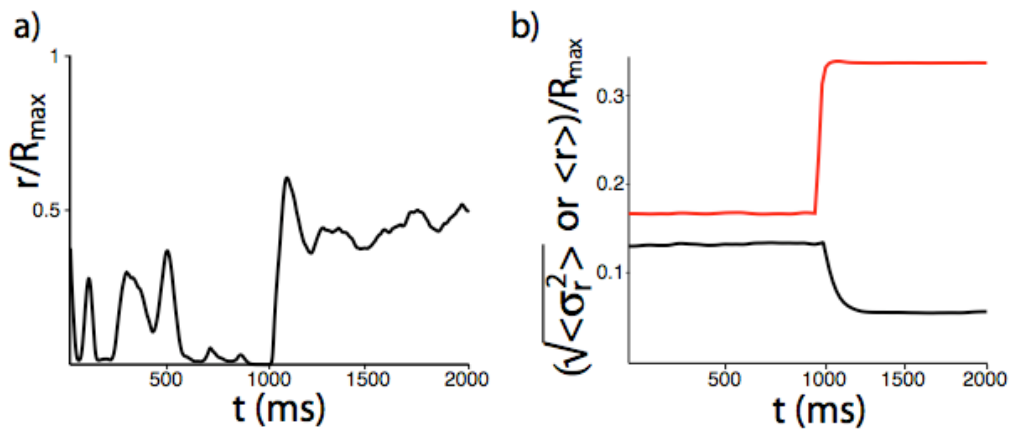

Figure 3. Firing rates and response variability normalized by the maximum firing rate, $R_{max}$, for the same network, stimulus and noise as in Figure 2, but for a network with both spontaneous activity and injected noise. a) The response of a typical neuron. b) The average firing rate (red trace) increases and the response variability (black trace) decreases when the stimulus input is present, but it does not go to zero as in Figure 2b.

## Response to Periodic Input

For Figures 2 & 3, the stimulus consisted of a step input applied identically to all neurons. To investigate the effect of more interesting and realistic stimuli on the chaotic activity of a recurrent network, we consider inputs with non-homogeneous spatio-temporal structure. Specifically, we introduce inputs that oscillate in a sinusoidal manner with amplitude $I$ and frequency $f$ and examine how the suppression of fluctuations depends on their amplitude and frequency. In many cases, neurons in a local population have diverse stimulus selectivities, so a particular stimulus may induce little change in the total activity across the network. To mimic this situation, we give these oscillating inputs a different phase for each neuron (in terms of a visual stimulus, this is equivalent to presenting a stationary, counterphase grating to a population of simple cells with different spatial-phase selectivities). Specifically, $I_i = I\cos(2\pi ft + \theta_i)$, where $\theta_i$ is chosen randomly from a uniform distribution between 0 and $2\pi$. The randomly assigned phases ensure that the spatial pattern of input in our model network is not correlated with the pattern of recurrent connectivity.

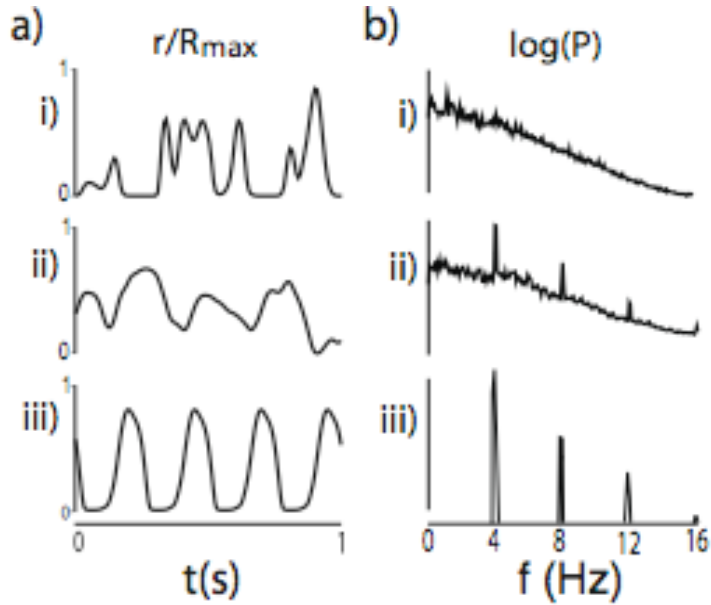

Figure 4. A chaotic network of 1000 neurons receiving sinusoidal 5 Hz input. a) Firing rates of typical network neurons (normalized by $R_{max}$). b) The logarithm of the power spectrum of the activity across the network. i) With no input ($I = 0$), network activity is chaotic. ii) In the presence of a weak input ($I/I_{1/2} = 0.1$), an oscillatory response is superposed on chaotic fluctuations. iii) For a stronger input ($I/I_{1/2} = 0.5$), the network response is periodic.

In the absence of a stimulus input, the firing rates of individual neurons fluctuate irregularly, as seen in Figure 2 (Figure 4a-i), and the power spectrum across network neurons is continuous and decays exponentially as a function of frequency (Figure 4b-i), a characteristic features of the chaotic state of this network (Sompolinsky, Crisanti, and Sommers, 1988). When the network is driven by a weak oscillatory input, the single-neuron response is a superposition of a periodic pattern induced by the input and a chaotic background (Figure 4a-ii). The power spectrum shows a continuous component due to the residual chaos and peaks at the frequency of the input and its harmonics, reflecting the periodic but non-sinusoidal component of the response(Figure 4b-ii). For an input with a larger amplitude, the firing rates of network neurons are periodic (Figure 4a-iii), and the power spectrum shows only peaks at the input frequency and its harmonics, with no continuous spectrum (Figure 4b-iii). This indicates a complete suppression of the internally generated fluctuations as in Figure 2a & b.

We have used a mean-field approach similar to that developed by Sompolinsky, Crisanti, and Sommers (1988) to analyze properties of the transition between chaotic a periodic responses to a periodic stimulus (Rajan, Abbott and Sompolinsky, 2009). This extends previous work on the effect of input on chaotic network activity (Molgedey, Schuchhardt and Schuster, 1992; Bertchinger and Natschläger, 2004) to continuous time models and periodic inputs. We find that there is a critical input intensity (a critical value of $I$) that depends on $f$ and $g$, below which network activity is chaotic though driven by the input (as in Figures 4a-ii & 4b-ii) and above which it is periodic (as in Figures 4a-iii & 4b-iii). A surprising feature of this critical amplitude is that it is a non-monotonic function of the frequency $f$ of the input. As a result, there is a "best" frequency at which it is easiest to entrain the network and suppress chaos. For the parameters we use, the "best" frequency is around 5 Hz, a frequency were many sensory systems tend to operate, and there are some initial experimental indications that this is indeed the optimal frequency for suppressing background activity by visual stimulation (White and Fiser, 2008). It is interesting that a preferred input frequency for entrainment arises even though the power spectrum of the spontaneous activity does not show any resonant features (Figure 4b-i).

**Principal Component Analysis of Spontaneous and Evoked Activity**

The results of the previous two sections revealed a regime in which an input generates a non-chaotic network response, even though the network is chaotic in the absence of input. Although the chaotic intrinsic activity has been completely suppressed in this network state, its imprint can still be detected in the spatial pattern of the non-chaotic activity.

The network state at any instant can be described by a point in an $N$-dimensional space with coordinates equal to the firing rates of the $N$ neurons. Over time, activity traverses a trajectory in this $N$-dimensional space. Principal component analysis can be used to delineate the subspace in which this trajectory predominantly lies. The analysis is done by diagonalizing the equal-time cross-correlation matrix of network firing rates, $<r_i(t)r_j(t)>$, where the angle brackets denote an average over time  The eigenvalues of this matrix expressed as a fraction of their sum (denoted by $\tilde{\lambda}_a$), indicate the distribution of variances across different orthogonal directions in the activity trajectory. In the spontaneous state, there are a number of significant contributors to the total variance, as indicated in Figure 5a. For this value of $g$, the leading 10% of the components account for 90% of the total variance. The variance associated with higher components falls off exponentially. It is interesting to note that the projections of the network activity onto the principal component directions fluctuate more rapidly for higher components (Figure 5c), revealing the interaction between the spatial and temporal structure of the chaotic fluctuations.

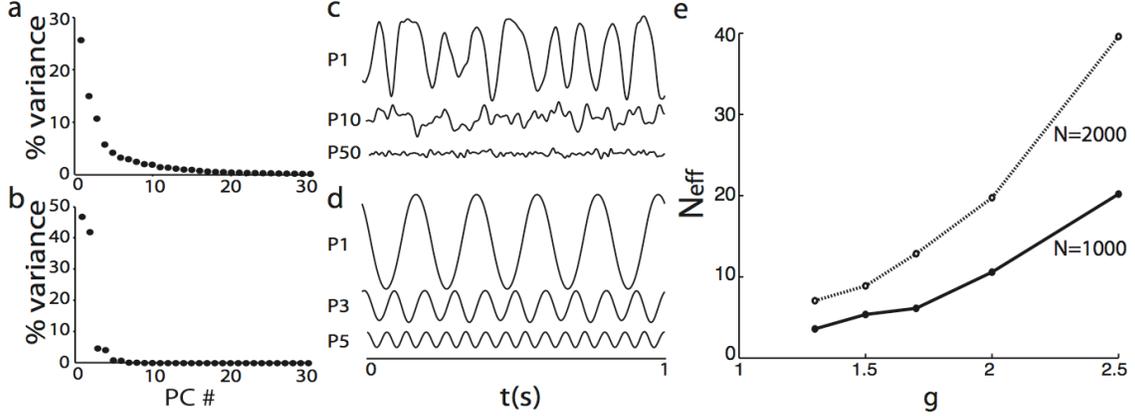

Figure 5: Principal component analysis of the chaotic spontaneous state and non-chaotic driven state. a) Percent variance accounted for by different principal components for chaotic spontaneous activity. b) Same as a, but for non-chaotic driven activity. c) Projections of the chaotic spontaneous activity onto principal component vectors 1, 10 and 50 (in decreasing order of variance). d) Projections of periodic driven activity onto principal components 1, 3, and 5. Projections onto components 2, 4, and 6 are similar except for being phase shifted by $\pi/2$. e) The effective dimension, $N_{\text{eff}}$, of the trajectory of chaotic spontaneous activity (defined in the text) as a function of $g$ for networks with 1000 (solid circles) or 2000 (open circles) neurons. Parameters: $g = 1.5$ for a-d, and $f = 5$ and $I/I_{1/2} = 0.7$ for b and d.

The non-chaotic driven state is approximately two dimensional (Figure 5b), with the two dimensions describing a circular oscillatory orbit. Projections of this orbit correspond to the oscillations $\pi/2$ apart in phase. The residual variance in the higher dimensions reflects higher harmonics arising from network nonlinearity, as illustrated by the projections in Figure 5d.

To quantify the dimension of the subspace containing the chaotic trajectory in more detail, we introduce the quantity

$$N_{\text{eff}} = \left( \sum_{a=1}^{N} \tilde{\lambda}_a^2 \right)^{-1}.$$

This provides a measure of the effective number of principal components describing a trajectory. For example, if $n$ principal components share the total variance equally, and the remaining $N - n$ principal components have zero variance, $N_{\text{eff}} = n$. For the chaotic spontaneous state in the networks we study, $N_{\text{eff}}$ increases with $g$ (Figure 5e), due to the higher amplitude and frequency content of the chaotic activity for large $g$. Note that $N_{\text{eff}}$ scales approximately with $N$, which means that large networks have proportionally higher-dimensional chaotic activity (compare the two traces in Figures 5e). The fact that the number of activated modes is only 2% of the system dimensionality, even for $g$ as high as 2.5, is another manifestation of the deterministic nature of the fluctuations. For comparison, we calculated $N_{\text{eff}}$ for a similar network driven by external white noise, with $g$ set below the chaotic transition at $g = 1$. In this case, $N_{\text{eff}}$ only assume such low values when $g$ is within a few percent of the critical value 1. The results in Figure 5 illustrate another feature of the suppression of spontaneous activity by input, which is that the PCA dimension $N_{\text{eff}}$ is reduced dramatically by the presence of the input.

**Network Effects on the Spatial Pattern of Evoked Activity**

In the non-chaotic regime, the temporal structure of network responses is largely determined by the input; they both oscillate at the same frequency, although the network activity includes harmonics not present in the input. The input does not, however, exert nearly as strong control on the spatial structure of the network response. The phases of the firing-rate oscillations of network neurons are only partially correlated with the phases of the inputs that drive them, and they are strongly influenced by the recurrent feedback.

We have seen that the orbit describing the activity in the non-chaotic driven state consists primarily of a circle in a two-dimensional subspace of the full $N$-dimensions describing neuronal activities. We now ask how this circle aligns relative to subspaces defined by different numbers of principal components that characterize the spontaneous activity. This relationship is difficult to visualize because both the chaotic subspace and the full space of network activities are high dimensional. To overcome this difficulty, we make use of the notion of "principal angles" between subspaces (Ipsen and Meyer, 1995).

The first principal angle is the angle between two unit vectors (called principal vectors), one in each subspace, that have the maximum overlap (dot product). Higher principal angles are defined recursively as the angles between pairs of unit vectors with the highest overlap that are orthogonal to the previously defined principal vectors. Specifically, for two subspaces of dimension $d_1$ and $d_2$ defined by the orthogonal unit vectors $\boldsymbol{V}_1^a$, for $a = 1, 2, ..., d_1$ and $\boldsymbol{V}_2^b$, for $b = 1, 2, ..., d_2$, the cosines of the principal angles are equal to the singular values of the $d_1$ by $d_2$ matrix formed from all the possible dot products of these two vectors. The resulting principal angles vary between 0 and $\pi/2$ with zero angles appearing when parts of the two subspaces overlap and $\pi/2$ corresponding to directions in which the two subspaces are completely non-overlapping. The angle between two subspaces is the largest of their principal angles. This definition is illustrated in Figure 6a where we show the irregular trajectory of the chaotic spontaneous activity, described by its two leading principal components (black curve in Figure 6a). The circular orbit of the periodic activity (red curve in Figure 6a) has been rotated by the smaller of its two principal angles. The angle between these two subspaces (the angle depicted in Figure 6a) is then the remaining angle through which the periodic orbit would have to be rotated to bring it into alignment with the horizontal plane containing the two-dimensional projection of the chaotic trajectory.

Figure 6a shows the angle between the subspaces defined by the first two principal components of the orbit of periodic driven activity and the first two principal components of the chaotic spontaneous activity. We now extend this idea to a comparison of the two-dimensional subspace of the periodic orbit and subspaces defined by the first $m$ principal components of the chaotic spontaneous activity. This allows us to see how the orbit lies in the full $N$-dimensional space of neuronal activities relative to the trajectory of the chaotic spontaneous activity. The results (Figure 6b, red dots) show that this angle is close to $\pi/2$ for small $m$, equivalent to the angle between two randomly chosen subspaces. However, the value drops quickly for subspaces defined by progressively more of the leading principal components of the chaotic activity. Ultimately, this angle approaches zero when all $N$ of the chaotic principal component vectors are considered, as it must, because these span the entire space of network activities.

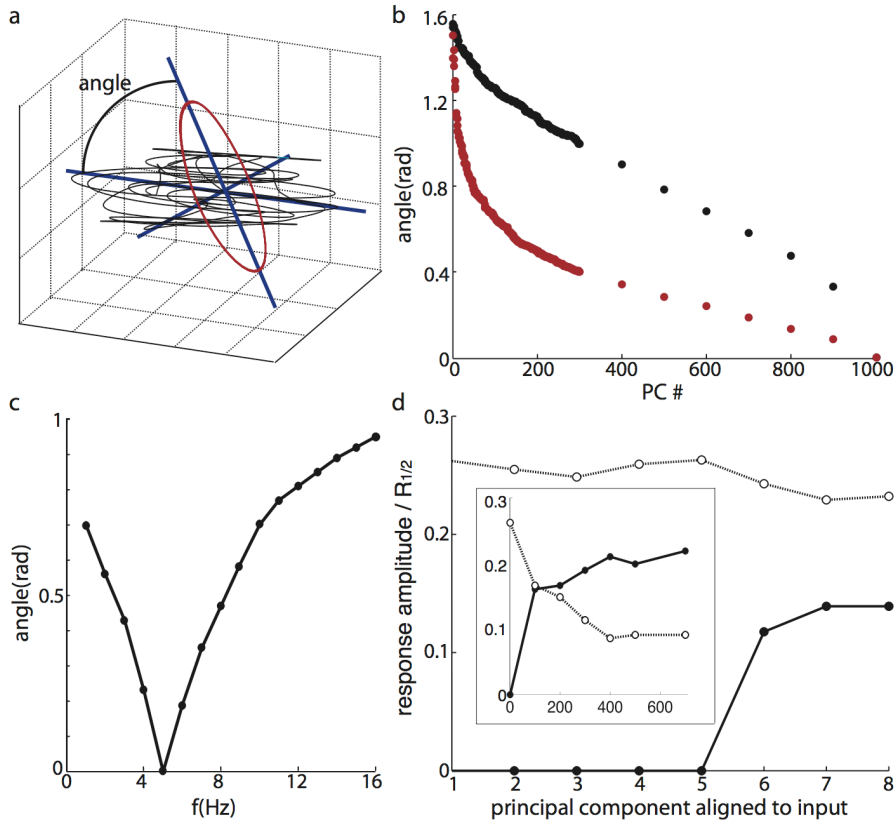

Figure 6: Spatial pattern of network responses. a) Definition of the angle between the subspace defined by the first two components of the chaotic activity (black curve) and a two-dimensional description of the periodic orbit (red curve). b) Relationship between the orientation of periodic and chaotic trajectories. Angles between the subspace defined by the two principal components of the non-chaotic driven state and subspaces formed by principal components 1 through $m$ of the chaotic spontaneous activity, where $m$ appears on the horizontal axis (red dots). Black dots show the analogous angles but with the two-dimensional subspace defined by random input phases replacing the subspace of the non-chaotic driven activity. c) Effect of input frequency on the orientation of the periodic orbit. The angle (vertical axis) between the subspaces defined by the two leading principal components of non-chaotic driven activity at different frequencies (horizontal axis) and these two vectors for a 5 Hz input frequency. d) Network selectivity to different spatial patterns of input. Signal (dashed curves and open circles) and noise (solid curves and filled circles) amplitudes in response to inputs aligned to the leading principal components of the spontaneous activity of the network. The inset shows a larger range on a coarser scale. Parameters: $I/I_{1/2} = 0.7$ and $f = 5$ Hz for b, $I/I_{1/2} = 1.0$ for c, and $I/I_{1/2} = 0.2$ and $f = 2$ Hz for d.

In the periodic state, the temporal phases of the different neurons determine the orientation of the orbit in the space of neuronal activities. The rapidly falling angle between this orbit and the subspaces defined by spatial patterns dominating the chaotic state (Figure 6b, red dots) indicates that these phases are strongly influenced by the recurrent connectivity that in turn determines the spatial pattern of the spontaneous activity. As an indication of the magnitude of this effect, we note that the angles between the random phase sinusoidal trajectory of the input to the network and the same chaotic subspaces are much larger than those associated with the periodic network activity (Figure 6b, black dots).

**Temporal Frequency Modulation of Spatial Patterns**

Although recurrent feedback in the network plays an important role in the spatial structure of driven network responses, the spatial pattern of the activity is not fixed but instead is shaped by a complex interaction between the driving input and the intrinsic network dynamics. It is therefore sensitive to both the amplitude and the frequency of this drive. To see this, we examine how the orientation of the approximately two-dimensional periodic orbit of driven network activity in the non-chaotic regime depends on input frequency. We use the technique of principal angles described in the previous section, to examine how the orientation of the oscillatory orbit changes when the input frequency is varied. For comparison purposes, we choose the dominant two-dimensional subspace of the network oscillatory responses to a driving input at 5 Hz as a reference. We then calculate the principal angles between this subspace and the corresponding subspaces evoked by inputs with different frequencies. The result shown in Figure 6c indicates that the orientation of the orbit for these driven states rotates as the input frequency changes.
    The frequency dependence of the orientation of the evoked response is likely related to the effect seen in Figure 6c in which higher frequency activity is projected onto higher principal components of the spontaneous activity. This causes the orbit of driven activity to rotate in the direction of higher-order principal components of the spontaneous activity as the stimulus frequency increases. In addition, the larger the stimulus amplitude, the closer the response phases of the neurons will be to the random phases of their external inputs (results not shown).

**Network Selectivity**

We have shown that the response of a network to random-phase input is strongly affected by the spatial structure of spontaneous activity (Figure 6b). We now ask if the spatial patterns that dominate the spontaneous activity in a network correspond to the spatial input patterns to which the network responds most vigorously. Rather than using random-phase inputs, we now aligned the inputs to our network along the directions defined by different principal components of its spontaneous activity. Specifically, the input to neuron $i$ is set to $IV_i^a\cos(2\pi ft)$, where $I$ is the amplitude factor and $V_i^a$ is the $i^{th}$ component of principal component vector $a$ of the spontaneous activity. The index $a$ is ordered so that $a = 1$ corresponds to the principal component with largest variance and $a = N$ the least. To analyze the results of using this input, we divide the response into a signal component corresponding to the trial-averaged response, and a noise component consisting of the fluctuations around this average response. We call the amplitude of the signal component of the response the "signal amplitude" and the standard deviation of the fluctuations the "noise amplitude".

As seen in Figure 6d the amplitude of the signal component of the response decreases slowly as a function of which principal component is used to define the input. A more dramatic effect is seen on the noise component of the response. For the input amplitude used in Figure 6d, inputs aligned to the first 5 principal components of the spontaneous activity completely suppress the chaotic noise, resulting in periodic driven activity. For higher-order principal components, the network activity is chaotic. Thus, the "noise" shows more sensitivity to the spatial structure of the input than the signal.

**Discussion**

Our results suggest that experiments that study the stimulus-dependence of the typically ignored noise component of responses should be interesting and could provide insight into the nature and origin of activity fluctuations. Response variability and ongoing activity is sometimes modeled as arising from a stochastic process external to the network generating the responses. This stochastic noise is then added linearly to the signal to create the total neuronal activity in the evoked state. Our results indicate that recurrent dynamics of the cortical circuit is likely to contribute significantly to the emergence of irregular neuronal activity, and that the interaction between such deterministic "noise" and external drive is highly nonlinear. In our work (Rajan, Abbott and Sompolinsky, 2009), we have shown that the stimulus causes a strong suppression of activity fluctuations and furthermore that the nonlinear interaction between the relatively slow chaotic fluctuations and the stimulus results in a non-monotonic frequency dependence of the noise suppression.

An important feature of the networks we study is that the variance of the synaptic strengths across the network controls the emergence of interesting complex dynamics. This has important implications for experiments because it suggests that the most interesting and relevant modulators of networks may be substances or activity-dependent modulations that do not necessarily change properties of synapses on average, but rather change synaptic variance. Synaptic variance can be changed either by modifying the range over which synaptic strengths vary across a population of synapses, as we have done here, or by modifying the release probability and variability of quantal size at single synapses. Such modulators might be viewed as less significant because they do not change the net balance between excitation and inhibition. However, network modeling suggests that such modulations are of great importance in controlling the state of the neuronal circuit.

The random character of the connectivity in our network precludes a simple description of the spatial activity patterns in terms of topographically organized maps. Our analysis shows that even in cortical areas where the underlying connectivity does not exhibit systematic topography, dissecting the spatial patterns of fluctuations in neuronal activity can reveal important insight about both intrinsic network dynamics and stimulus selectivity. Principal component analysis revealed that despite the fact that the network connectivity matrix is full rank, the effective dimensionality of the chaotic fluctuations is much smaller than the number of neurons in the network. This suppression of spatial modes is much stronger than expected from a linear network low-pass filtering a spatio-temporal white noise input. Furthermore, as in the temporal domain, active spatial patterns exhibit strong nonlinear interaction between external driving inputs and intrinsic dynamics. Surprisingly, even when the stimulus amplitude is strong enough to fully entrain the temporal pattern of network activity, spatial organization of the activity is still strongly influenced by recurrent dynamics, as shown in Figures 6c and 6d.

We have presented tools for analyzing the spatial structure of chaotic and non-chaotic population responses based on principal component analysis and angles between the resulting subspaces. Principal component analysis has, been applied profitably to neuronal recordings (see, for example, Broome, Jayaraman and Laurent, 2006). These analyses often plot activity trajectories corresponding to different network states using the fixed principal component coordinates derived from combined activities under all conditions. Our analysis offers a complementary approach whereby principal components are derived for each stimulus condition separately, and principal angles are used to reveal not only the difference between the shapes of trajectories corresponding to different network states, but also the difference in the orientation of the low dimensional subspaces of these trajectories within the full space of neuronal activity.

Many models of selectivity in cortical circuits rely on knowledge of the spatial organization of afferent inputs as well as cortical connectivity. However, in many cortical areas, such information is not available. Our results show that experimentally accessible spatial patterns of spontaneous activity (e.g. from voltage- or calcium-sensitive optical imaging experiments) can be used to infer the stimulus selectivity induced by the network dynamics and to design spatially extended stimuli that evoke strong responses. This is particularly true when selectivity is measured in terms of the ability of a stimulus to entrain the neural dynamics, as in Figure 6d. In general, our results indicate that the analysis of spontaneous activity can provide valuable information about the computational implications of neuronal circuitry.

**Acknowledgments**

Research of KR and LA supported by National Science Foundation grant IBN-0235463 and an NIH Director's Pioneer Award, part of the NIH Roadmap for Medical Research, through grant number 5-DP1-OD114-02. HS is partially supported by grants from the Israel Science Foundation and the McDonnell

Foundation. This research was also supported by the Swartz Foundation through the Swartz Centers at Columbia and Harvard Universities.